\documentclass[usegraphicx,usenatbib]{mn2e}

\def\ni{\noindent}

\def\rysunek#1#2#3{
\begin{figure}
\includegraphics[width=88mm]{#1}
\caption{#2}
\label{#3}
\end{figure}
}

\title{THE END OF THE DARK AGES IN MOND}

\author[S.~Stachniewicz, M.~Kutschera]
{S.~Stachniewicz$^1$, M.~Kutschera$^{1,2}$\\
$^1$Astrophysics Division, H.Niewodnicza\'nski Institute of Nuclear Physics,
ul. Radzikowskiego 152,\\
31-342 Krak\'ow, Poland\\
$^2$Institute of Physics, Jagiellonian University, ul. Reymonta 4,\\
30-059 Krak\'ow, Poland
}
\date{}

\begin{document}
\maketitle

\begin{abstract}
We study the evolution of a spherically symmetric density perturbation
in the Modified Newtonian Dynamics (MOND) model applied to the net
acceleration over Hubble flow. The background cosmological
model is a $\Lambda$-dominated, low-$\Omega_b$ Friedmann model with no Cold
Dark Matter. We include thermal processes and non-equilibrium chemical
evolution of the collapsing gas. We find that under these assumptions the first
low-mass objects ($M \le 3\times 10^4 M_{\odot}$) may collapse already for
$z\sim 30$, which is in quite good agreement with the recent WMAP results.
A lower value of $a_0$ would lead to much slower collapse of such objects.
\end{abstract}

\begin{keywords}
hydrodynamics -- gravitation -- instabilities -- dark matter -- early Universe.
\end{keywords}

\section{Introduction}

Recent developments in cosmological observations have led to the so-called
cosmological concordance model with $\Omega_b$ around 0.03, $\Omega_m$
(dark+baryonic) around 0.3 and $\Omega_{\Lambda}$ around 0.7. However,
as the $\Lambda$CDM models are dominated by hypothetical vacuum energy
and non-baryonic dark matter contributions, some scientists are looking for
different solutions.
Perhaps the most intereresting alternative model is the Modified
Newtonian Dynamics model (MOND) proposed by \citet{Mil83}.
It assumes that there is no non-baryonic dark matter (or it is negligible)
and the apparent lack of matter is only due to the modification of
dynamics or gravity for small accelerations ($a\ll a_0$ where $a_0$ is
some constant). This model seems to work very well for spiral galaxies
and many other types of objects \citep{SMG02}. However, it carries with it
some unresolved problems (e.g. lack of covariance).

The problem of the structure formation in MOND was also explored by
\citet{San01} and \citet{Nus02},
but they were interested in large cosmic structures and did
not include gas effects. It is a good approximation for large scales, but gas
effects play a crucial role in the small-scale structure formation.

\citet{Sta01b} studied possible implications of the MOND model for the
formation of the very first objects in the Universe. However, there we had
applied MOND to total gravitational acceleration, and this approach can lead
to some paradoxes, e.g. time and the very occurence of collapse hardly depend
on the initial overdensity. In this paper, we attempt a different approach;
namely, we apply MOND to the {\bf net} acceleration over the Hubble flow only.

\section{MOND vs standard theory of linear perturbations}

If one wants to apply the MOND model to structure formation calculations, they
encounter a number of difficulties. First of all, MOND is not a
theory, but rather a phenomenological model. In its present form,
MOND is inconsistent with General Relativity. Although \citet{Bek04} found
a relativistic gravitation theory that gives MOND-like predictions in the
low-gravity limit, it needs to be verified.

MOND is a model that modifies either dynamics or gravity (in this paper we
assume the latter). It introduces a new fundamental scale, usually
called $a_0$. Gravitational fields much stronger than $a_0$ are identical to
their Newtonian limit $g_N$, and very weak fields are $\sqrt{a_0 g_N}$.
According to \citet{SVe98}, the value of the fundamental
acceleration scale is $a_0 = 1.2 \times 10^{-8} \rm{cm}/\rm{s}^2$. More
precisely, the strength of the gravitational field may be written as

\begin{equation}
\mu\left({g \over a_0}\right) \vec{g} = \vec{g}_N ,
\end{equation}

\ni where $\mu(x)$ is some function that interpolates between these
two extreme cases. This function is not specified in the model.
We have decided to apply the function used by \citet{SVe98}:

\begin{equation}
\mu(x) = {x \over \sqrt{1+x^2}}
\end{equation}

\ni and, finally,

\begin{equation}
\vec{g} = \vec{g}_N \sqrt{1+\sqrt{1+({2 \over x})^2} \over 2}
\end{equation}

\ni where $x = g_N/a_0$.

The conseqences of MOND for cosmology have not been studied in detail yet.
\citet{San98} suggested that because in the early Universe
the MOND radius is much lower than that of the horizon, the evolution
of the scale factor is described by the standard Friedmann equations.
Also in theory described by \citet{Bek04} cosmological models are similar
to the Friedmann ones so we follow this assumption and study the formation of
the first objects in the Universe with modified dynamics.

We adopted this approach in \citet{Sta01b}, but, as mentioned in the
Introduction, this may lead to some paradoxes -- in particular, even underdense
regions of the Universe may collapse. In order to avoid that problem, we have
tried another
approach appliying MOND to the {\bf net} acceleration/deceleration over the
Hubble flow. In this approach, only clouds with {\bf positive} overdensity
may stop their expansion and recollapse, so the final formula for gravitational
acceleration (or deceleration) is

\begin{equation}
\mu\left({|\vec{g} - \vec{g_H}| \over a_0}\right) (\vec{g} - \vec{g_H}) =
\vec{g}_N - \vec{g_H} ,
\end{equation}

\ni where $g_H$ is the deceleration (or acceleration) due to the Hubble flow.

\subsection{Collapse of perfect gas}

If we take perfect gas instead of the pressureless fluid, the evolution will
look different because the effects of pressure will moderate the recollapse,
especially for small systems. Since we assume spherical symmetry, Lagrangian
coordinates are used.

In the Newtonian case, dynamics is governed by the following equations:

\begin{eqnarray}
{dM \over dr} & = & 4\pi r^2 \varrho, \label{ciaglosc}\\
{dr \over dt} & = & v , \label{promien}\\
{dv \over dt} & = & -4\pi r^2 {dp \over dM}-{GM(r) \over r^2} ,
\label{predkosc}\\
{du \over dt} & = & {p \over \varrho^2} {d \varrho \over dt} + {\Lambda \over
\varrho} ,
\label{energia}
\end{eqnarray}

\ni where $r$ is the radius of a sphere of mass $M$, $u$ is the internal
energy per unit mass, $p$ is pressure and $\varrho$ is mass
density. Here, Eq.(\ref{ciaglosc}) is the continuity equation, (\ref{promien})
and (\ref{predkosc}) give acceleration and (\ref{energia}) accounts for
energy conservation.
The last term in Eq.(\ref{energia}) describes gas cooling/heating, with
$\Lambda$ being energy absorption (emission) rate per
unit volume, given in detail in \citet{Sta01a}.

We use the equation of state of perfect gas

\begin{equation} p= (\gamma -1) \varrho u , \end{equation}

\ni where $\gamma = 5/3$, as the primordial baryonic matter after
recombination
is assumed to be composed mainly of monoatomic hydrogen and helium, with
the fraction of molecular hydrogen $H_2$ always less than $10^{-3}$.

In the case of modified gravity, equation (\ref{predkosc}) will look somewhat
different:

\begin{equation}
{dv \over dt}= -4\pi r^2 {dp \over dM}-g_H-a_0f\left({GM(r) \over a_0 r^2}-{g_H \over a_0}\right) ,
\label{modpr}
\end{equation}

\ni where $f(x)$ is inverse to function $\mu(x)$ mentioned
before, asymptotically equal to $x$ for $x\gg a_0$ and to $\sqrt{a_0x}$
for $x\ll a_0$, while $g_H$ may be expressed as

\begin{equation}
g_H = {1 \over 2} {H_0}^2 \left[(z+1)^3\Omega_b + 2 \left( (z+1)^4\Omega_r -
\Omega_{\Lambda}\right) \right] r .
\label{gH}
\end{equation}

\ni Here, $H_0$ is the current value of the Hubble parameter, $\Omega_b$,
$\Omega_r$ and $\Omega_{\Lambda}$ are the current fractions
of baryons, radiation and dark energy in terms of the critical density of the
Universe, and $z$ is the redshift.

\subsection{Chemical reactions and thermal effects}

In our calculations, we include all of the relevant thermal and chemical
processes in
the primordial gas. In this paper, we have taken into account nine species:
H, H$^-$, H$^+$, He, He$^+$, He$^{++}$, H$_2$, H$_2^+$ and e$^-$.
The abundance of various species changes with time due to chemical reactions,
ionization and dissociation photoprocesses. The chemical reactions
include such processes as ionization of hydrogen and helium
by electrons, recombination of ions with electrons,
formation of negative hydrogen ions, formation of H$_2$
molecules, etc. A full list of the relevant chemical reactions and appriopriate
formulae is given in \citet{Sta01a}.

The time evolution of the number density of component $n_i$ is
described by the kinetic equation:

\begin{equation}
{dn_i \over dt} = \sum_{l=1}^9 \sum_{m=1}^9 a_{lmi} k_{lm} n_l n_m +
\sum_{j=1}^9 b_{ji} \kappa_j n_j .
\label{chemia}
\end{equation}

\ni The first component on the right-hand side of this equation describes
the chemical reactions, and the other one accounts for photoionization and
photodissociation processes. Coefficients
$k_{lm}$ are reaction rates, quantities $\kappa_n$ are photoionization
or photodissociation
rates, and $a_{lmi}$ and $b_{ji}$ are numbers equal to 0, $\pm 1$ or $\pm 2$
depending on the reaction.
All reaction rates, as well as photoionization and photodissociation rates
are given in \citet{Sta01a}.

The cooling (heating) function $\Lambda_{cool}$ includes effects of
collisional ionization of H, He and He$^+$, recombination to H, He and He$^+$,
collisional excitation of H and He$^+$, Bremsstrahlung, Compton cooling
and cooling by deexcitation of H$_2$ molecules.
The formulae for the heating/cooling contributions of various processes
are given in \citet{Sta01a}.

\section{Code used in the simulations and initial conditions}

In the simulations we used the code described in \citet{Sta01a}, based on
those presented by \citet{Tho95} and \citet{Hai96}.
This is a standard, one-dimensional, second-order
accurate, Lagrangian finite-difference scheme. The only changes consist in the
modification of gravity, with the dark matter fraction $\Omega_{dm}$
set to zero. However, it was necessary to make significant changes
in the initial conditions.

First of all, we started our calculations at the end of the
radiation-dominated era. For $\Omega_b=\Omega_m=0.02/h^2$, $z_{eq}=485$
as given by the formula provided by \citet{HE98},
$z_{eq}=2.50 \times 10^4 \Omega_0 h^2 \Theta_{2.7}^{-4}$,
where $\Theta_{2.7}=T_\gamma/2.7 K$, assuming $h=0.72$ and
$T_\gamma$=2.7277 K. We assumed that, like in the standard
cosmology, initial overdensities may grow only in the matter-dominated
era. We used our own code to calculate the initial chemical composition
and initial gas temperature. When compared with the results obtained by
\citet{Gal98}, ours are very similar -- the agreement is at
the level of 10-20\%.
The difference is probably due to the fact that Galli and Palla
had included more species (e.g. deuterium and lithium), and some reaction
rates they used had been somewhat different.

It is necessary to specify the initial baryonic matter perturbations, $\delta$.
We have decided to take values of the order of the contribution to the baryonic
matter power spectrum at a given scale and redshift. The problem is that at
these scales the power spectrum is known very roughly, as there are no
observational data and the relevant physical processes are not known very well.
The power spectrum may be calculated using the {\sc CMBFAST} program by
\citet{Sel96} $\Delta(k)=\sqrt{4\pi d^2_{norm} k^4 t_f(k,z)^2}$, where
$t_f$ is the transfer function, $k$ is the wavenumber and
the value of normalization constant $d^2_{norm}$ is obtained from
{\sc CMBFAST}. We also included the output of the CMB anisotropy program
written by N. Sugiyama, received from the author. The results are shown in
Fig. \ref{Delta}.

\rysunek{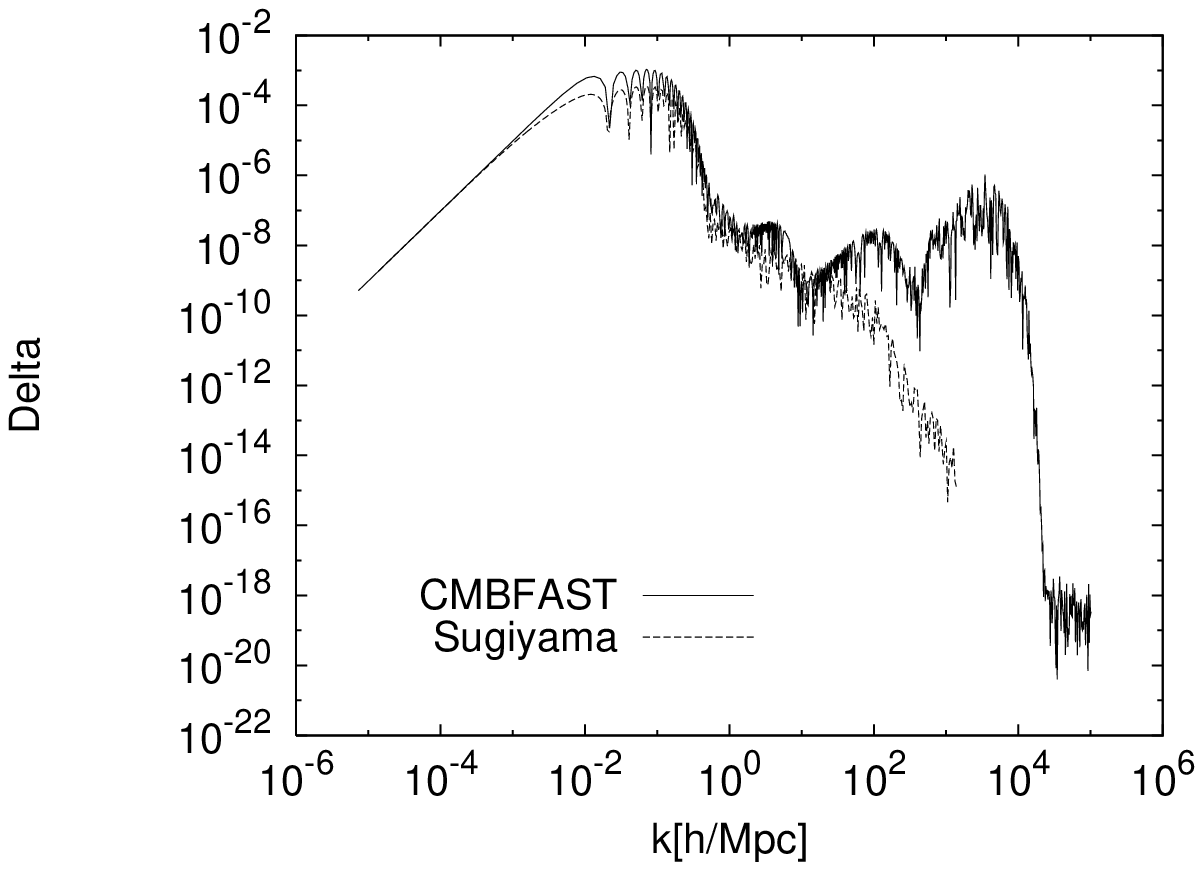}{Matter power spectrum.}{Delta}

The most interesting for us are the comoving scales around 10 kpc. For these
scales, both outputs differ significantly by a factor of 100. We decided to
take our initial overdensity from in between, i.e. equal to $10^{-9}$. As we
will show, these differences do not affect our results significantly.

We apply the initial density profiles in the form of a single spherical
Fourier mode, also used by \citet{Hai96}

\begin{equation} \varrho_b(r)=\Omega_b \varrho_c (1+\delta {\sin kr \over
kr})
,
\label{psinus} \end{equation}

\ni where $\varrho_c$ is the critical density of the Universe,
$\varrho_c=3H^2/8\pi G$, with $H$ being the actual value of the Hubble
parameter.

For this profile, there exist two distinguished radius values, namely
$R_0$ and $R_z$, which correspond to the first zero and the first minimum of
the function $\sin (kr)/kr$, respectively.
Inside the sphere of radius $R_0=\pi/k$ which contains mass
$M_0$, the local density contrast is positive. The
mass $M_0$ and the radius $R_0$ will be referred to as the cloud
mass and the cloud radius, respectively. The local
density contrast is negative for $R_z>r>R_0$, with average
density contrast vanishing for the sphere of radius $R_z=4.49341/k$ and
mass $M_z$. According to the gravitational instability
theory in the expanding Universe, the shell of radius $R_z$ will expand
together with the Hubble flow, not undergoing any additional deceleration.
This is why we regard this profile as very convenient in numerical simulations.
Namely, it eliminates the numerical edge effects and the mentioned shell simply
follows the Hubble expansion of the Universe. Thus, it can be regarded as the
perturbation boundary, with mass $M_z$ referred to as the bound mass.

It is worth noting that for radii not greater than $3/4 R_0$ this profile
is very similar to the Gaussian profile

\begin{equation} \varrho_i(r)=\Omega_i \varrho_c \left[ 1+\delta_i \exp \left(
{-r^2 \over 2R_f^2} \right) \right] \end{equation}

\ni with $R_f=1/2 R_0$.

As the initial velocity, we use the Hubble velocity:

\begin{equation}v(r)=Hr . \end{equation}

\section{Pressureless collapse in MOND}

We decided to describe the case of a pressureless collapse in MOND as this
makes it easier to understand why the results shown in the next section
hardly depend on the initial overdensities. It is much easier to trace shells 
since each shell may be treated separately, so it is enough to trace the 
evolution of a single shell with some initial overdensity and constant mass 
inside. Equation \ref{modpr} is reduced to

\begin{equation}
{dv \over dt}= -g_H-a_0f\left({GM(r) \over a_0 r^2}-{g_H \over a_0}\right) ,
\label{modprbezc}
\end{equation}

\ni where the expression for $g_H$ was shown in Eq. \ref{gH}. We wrote
a small program calculating the turnaround times for shells of various
masses and initial overdensities. It starts at the beginning of the
matter-dominated era with the initial velocity being the Hubble velocity
($v=Hr$). The results are shown in Fig. \ref{zturn}.

\rysunek{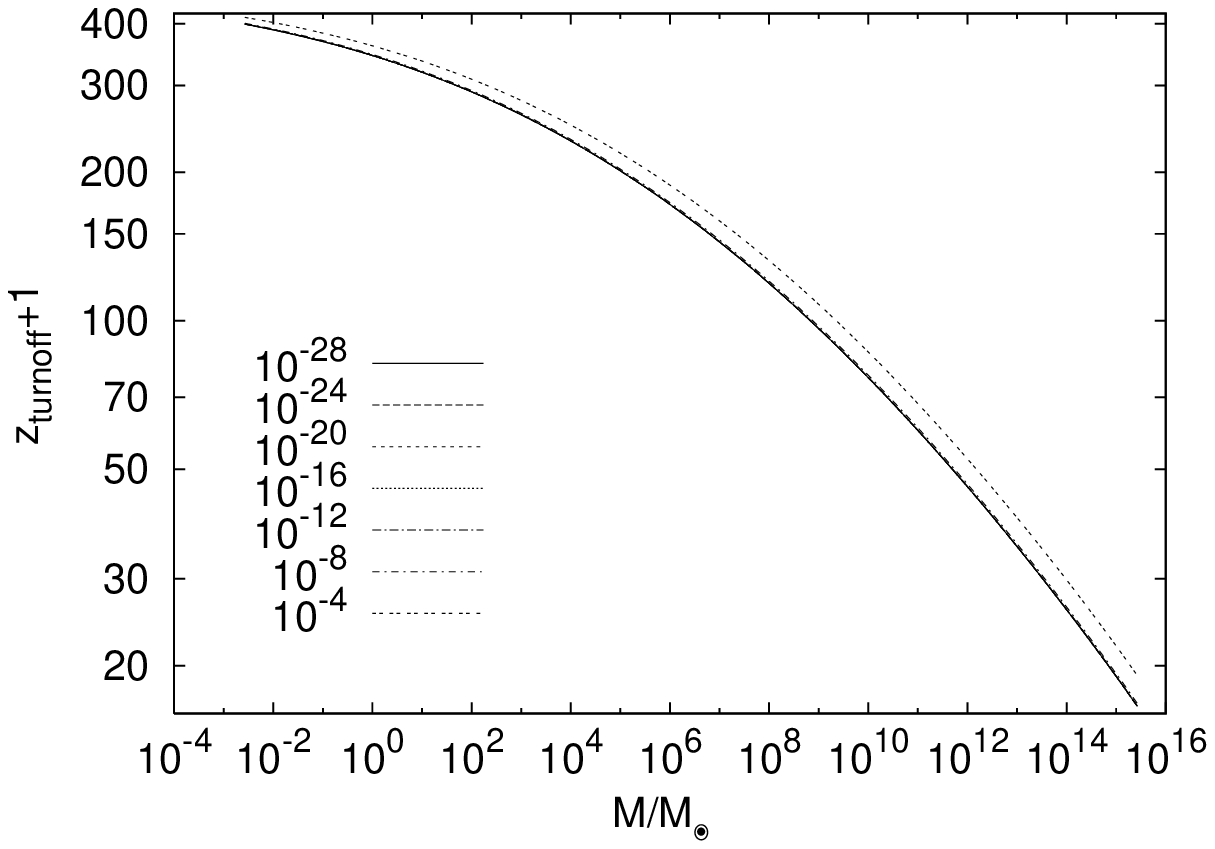}{Turnaround times for shells of various masses and initial
overdensities}{zturn}

As we can see, for initial overdensities of $10^{-8}$ and lower, the results
are almost indistinguishable and not very different from those for $10^{-4}$.
This looks very odd, but it reflects the fact that a smaller overdensity
means that the shell is very deep in the MOND regime and its behaviour is
strongly nonlinear. Now let us assume that a shell is always in the MOND
regime. For pure MOND Eq. \ref{modprbezc} takes the form

\begin{equation}
{dv \over dt}= -g_H-\sqrt{a_0\left({GM(r) \over r^2}-g_H\right)} .
\label{MONDbezc}
\end{equation}

It is a nonlinear equation, especially in the low overdensity limit.
This becomes obvious if we change variables. Instead of $r$, let us put
$(R-r)$, where $R$ is the radius corresponding to the mass M in a homogenous
Universe. In other words, $r$ is related to the actual overdensity $\delta$
and if it is much lower than $R$, it may be expressed as $\delta=3r/R$.
Given that, equation \ref{MONDbezc} may be written as

\begin{equation}
{d^2r \over dt^2}= \sqrt{3a_0G{M\over R^2}{r\over R}} .
\end{equation}

\ni Now we may express mass in terms of the cosmological parameters, redshift
and R, and we get

\begin{equation}
{d^2r \over dt^2}= \sqrt{{3\over 2}\Omega_ba_0H_0^2 (z+1)^3 r} .
\end{equation}

\ni Redshift $z$ is a function of time, and if we put
\hbox{$d\tau=[{3\over 2}\Omega_ba_0H_0^2 (z+1)^3]^{1/4} dt$}, we finally obtain

\begin{equation}
{d^2r \over d\tau^2}= \sqrt{r} .
\end{equation}

\section{Results}

We performed seventeen runs. The first eight were for the `standard' value
of $a_0$ ($1.2 \times 10^{-8} \mbox{cm/s$^2$}$), various masses of the cloud
($10^3 M_{\odot}$, $3\times 10^3 M_{\odot}$, $10^4 M_{\odot}$ and
$3\times 10^4 M_{\odot}$) and initial overdensities ($10^{-9}$ and $10^{-8}$,
the next three were for some lower value of $a_0$ ($1.2 \times 10^{-9}
\mbox{cm/s$^2$}$), initial overdensity $10^{-9}$ and masses of the cloud
equal to $3\times 10^3 M_{\odot}$, $10^4 M_{\odot}$ and
$3\times 10^4 M_{\odot}$, while the last one was for the `standard' value of
$a_0$ but with no H$_2$ cooling. The results are shown in Figs.
\ref{a0p01s1e3}-\ref{a0m13s3e4}.

\rysunek{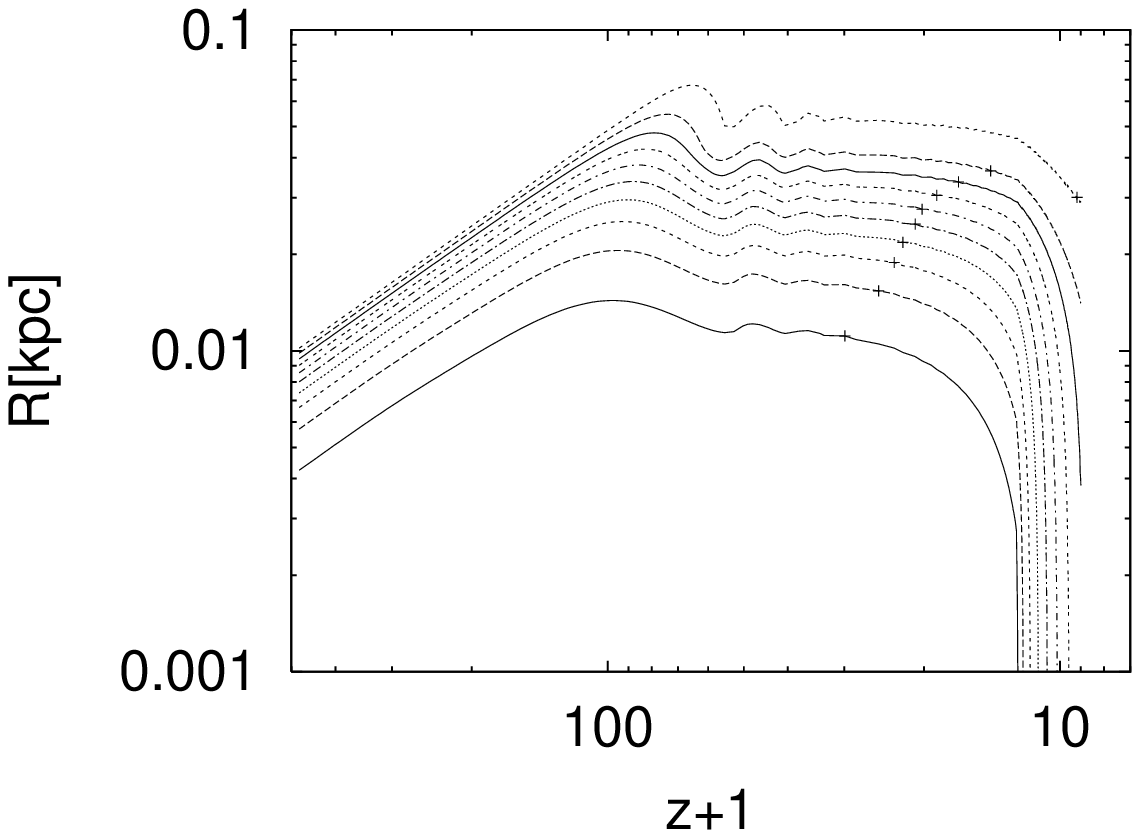}{Shell trajectories for the `standard' $a_0$,
$10^{-9}$ overdensity, $M=10^3 M_\odot$.}{a0p01s1e3}

\rysunek{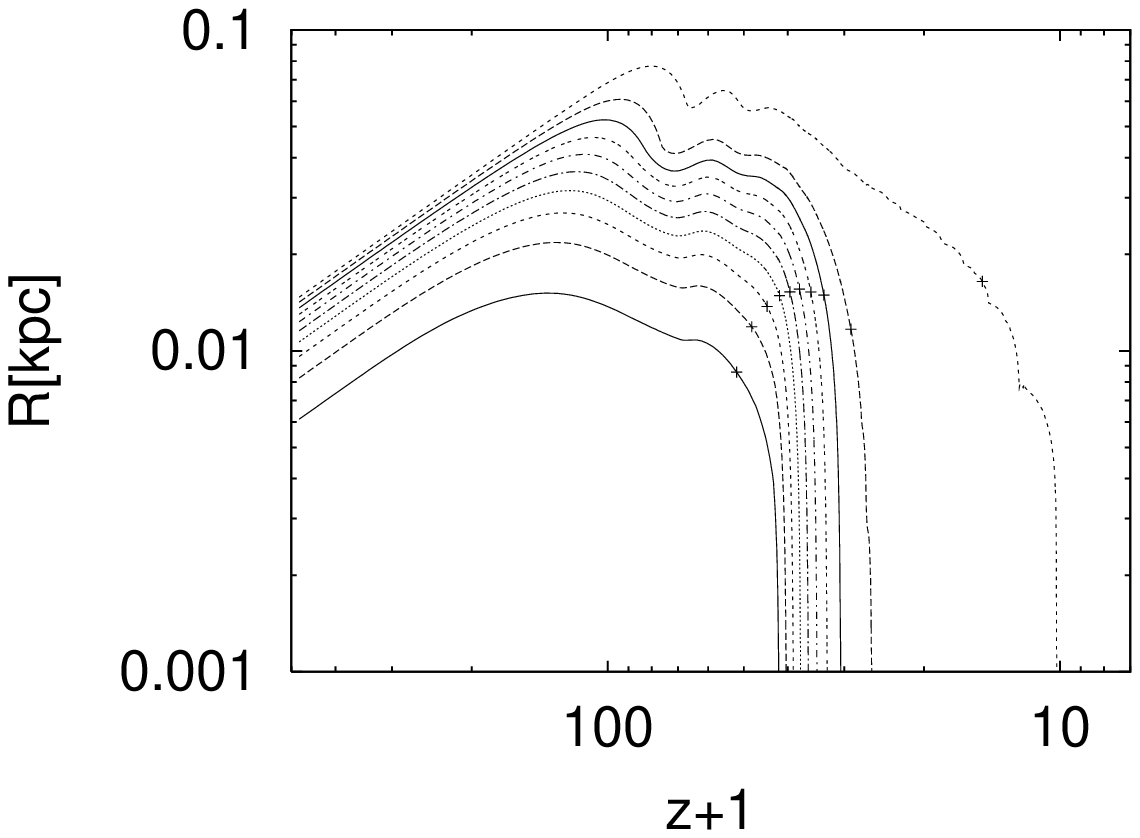}{Shell trajectories for the `standard' $a_0$,
$10^{-9}$ overdensity, $M=3\times 10^3 M_\odot$.}{a0p01s3e3}

\rysunek{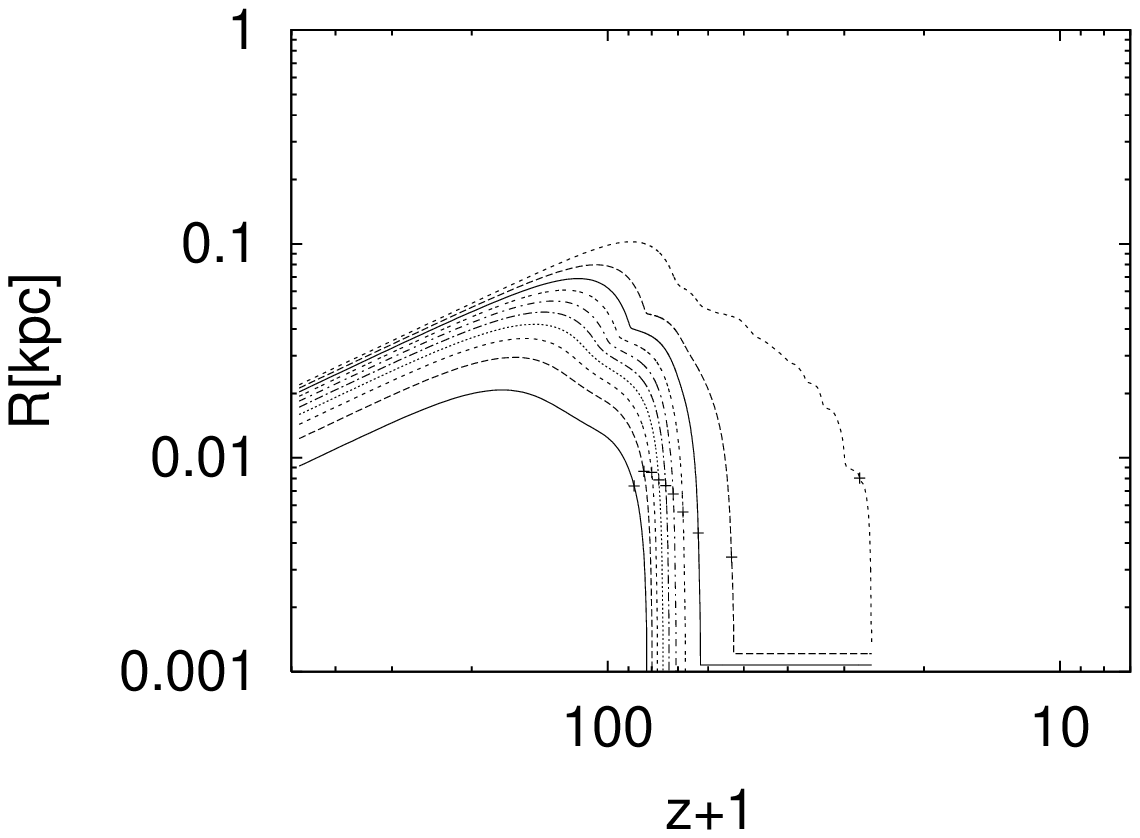}{Shell trajectories for the `standard' $a_0$,
$10^{-9}$ overdensity, $M=10^4 M_\odot$.}{a0p01s1e4}

\rysunek{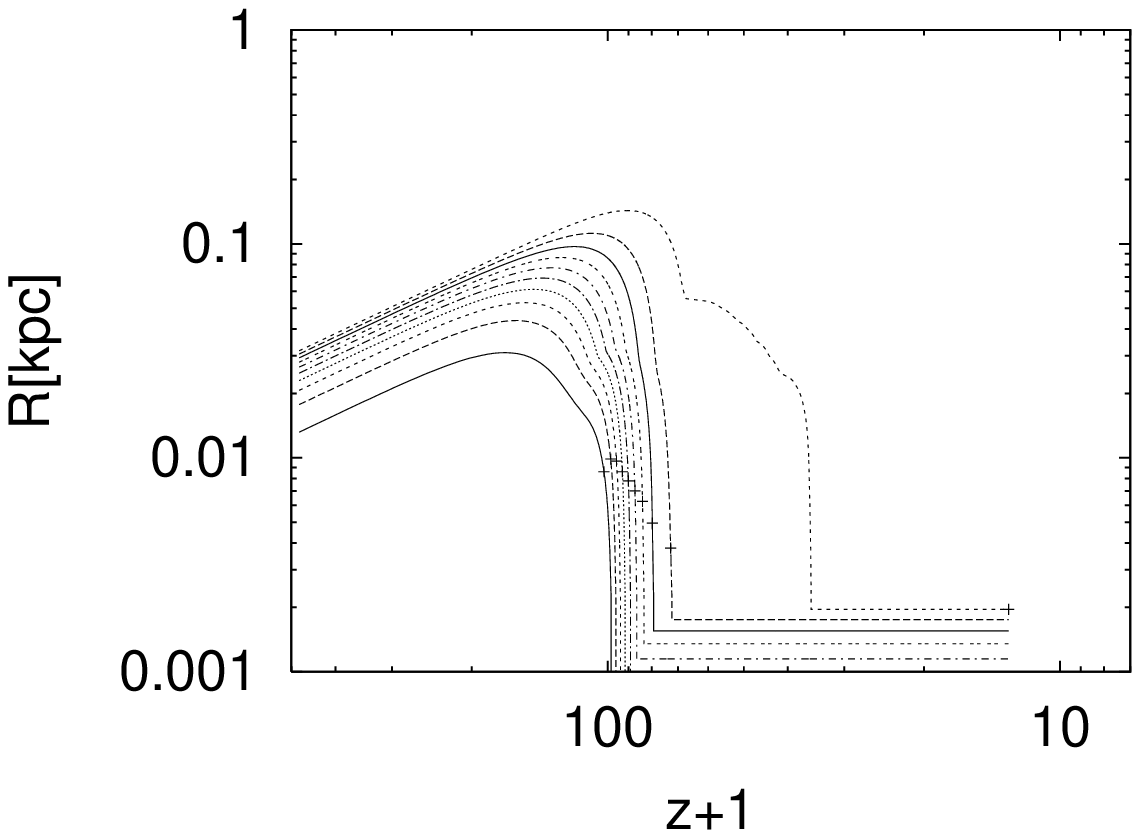}{Shell trajectories for the `standard' $a_0$,
$10^{-9}$ overdensity, $M=3\times 10^4 M_\odot$.}{a0p01s3e4}

\rysunek{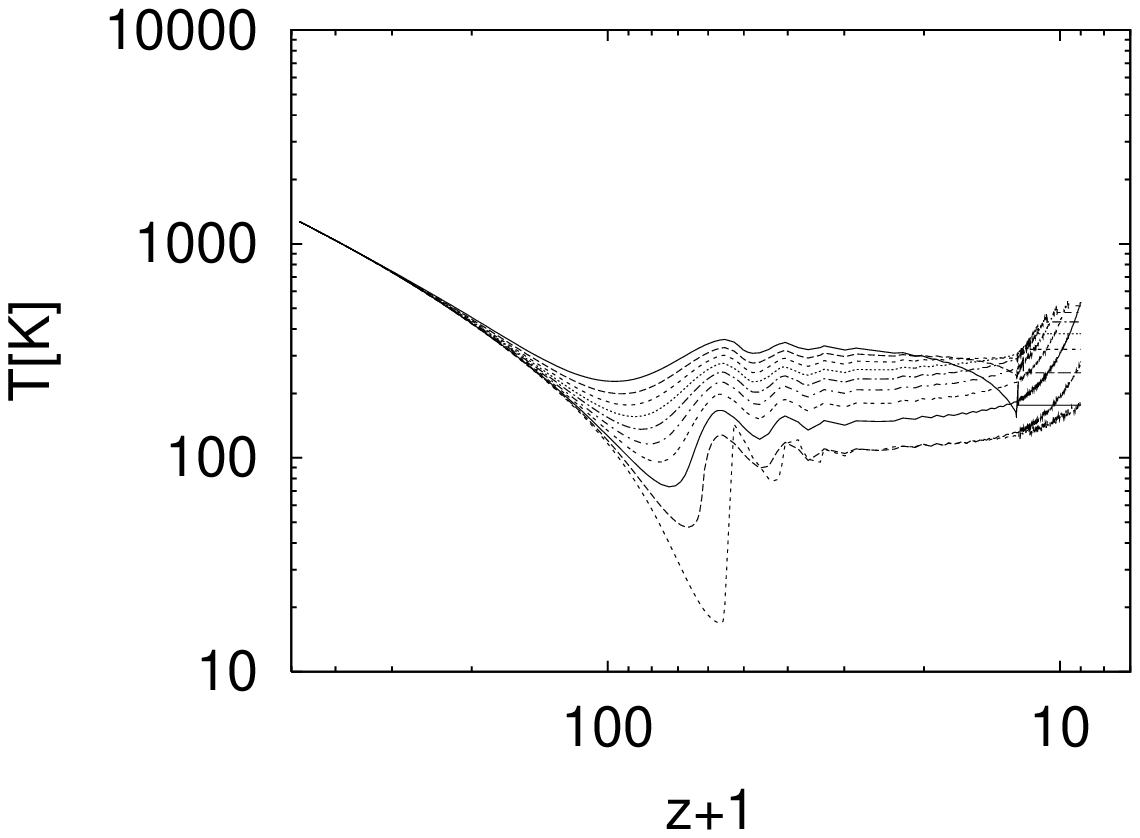}{Shell temperaturs for the `standard' $a_0$,
$10^{-9}$ overdensity, $M=10^3 M_\odot$.}{a0p01st}

\rysunek{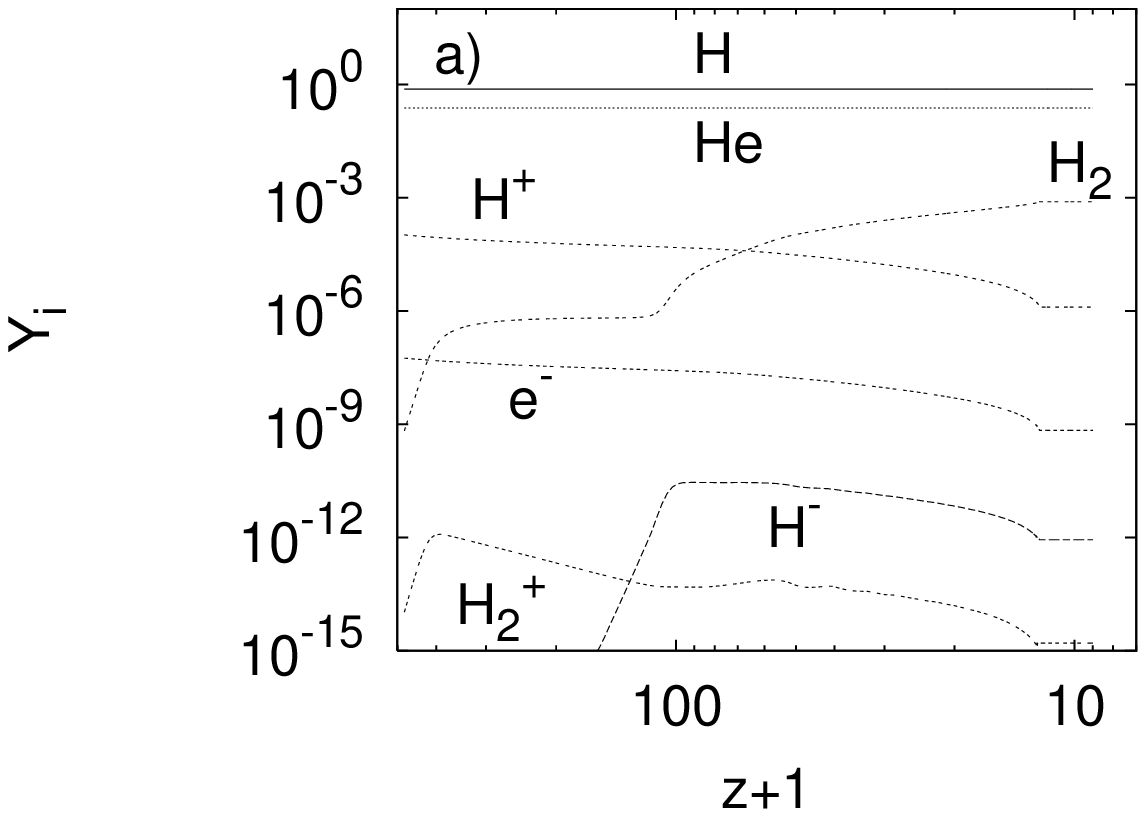}{Chemical evolution for the `standard' $a_0$,
$10^{-9}$ overdensity, $M=10^3 M_\odot$.}{a0p01sx}

\rysunek{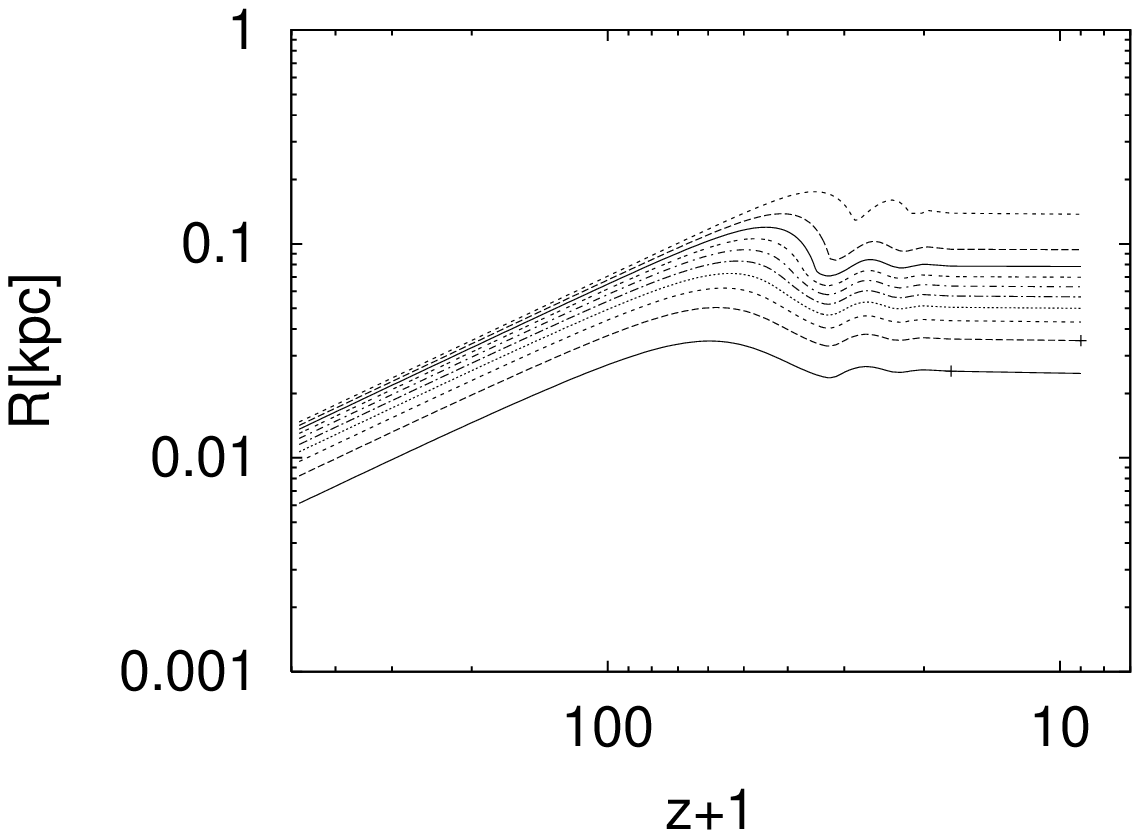}{Shell trajectories for the 'low' $a_0$, $10^{-9}$
overdensity, $M=3\times 10^3 M_\odot$.}{a0m13s3e3}

\rysunek{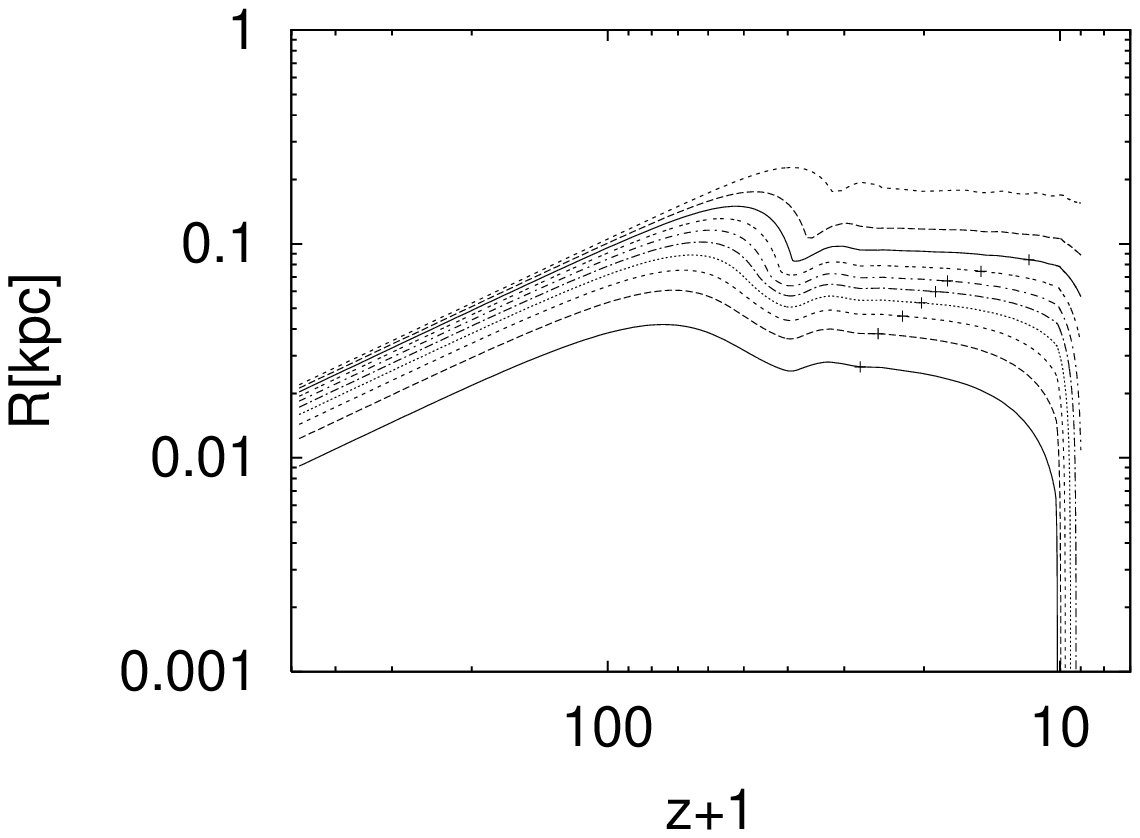}{Shell trajectories for the 'low' $a_0$, $10^{-9}$
overdensity, $M=10^4 M_\odot$.}{a0m13s1e4}

\rysunek{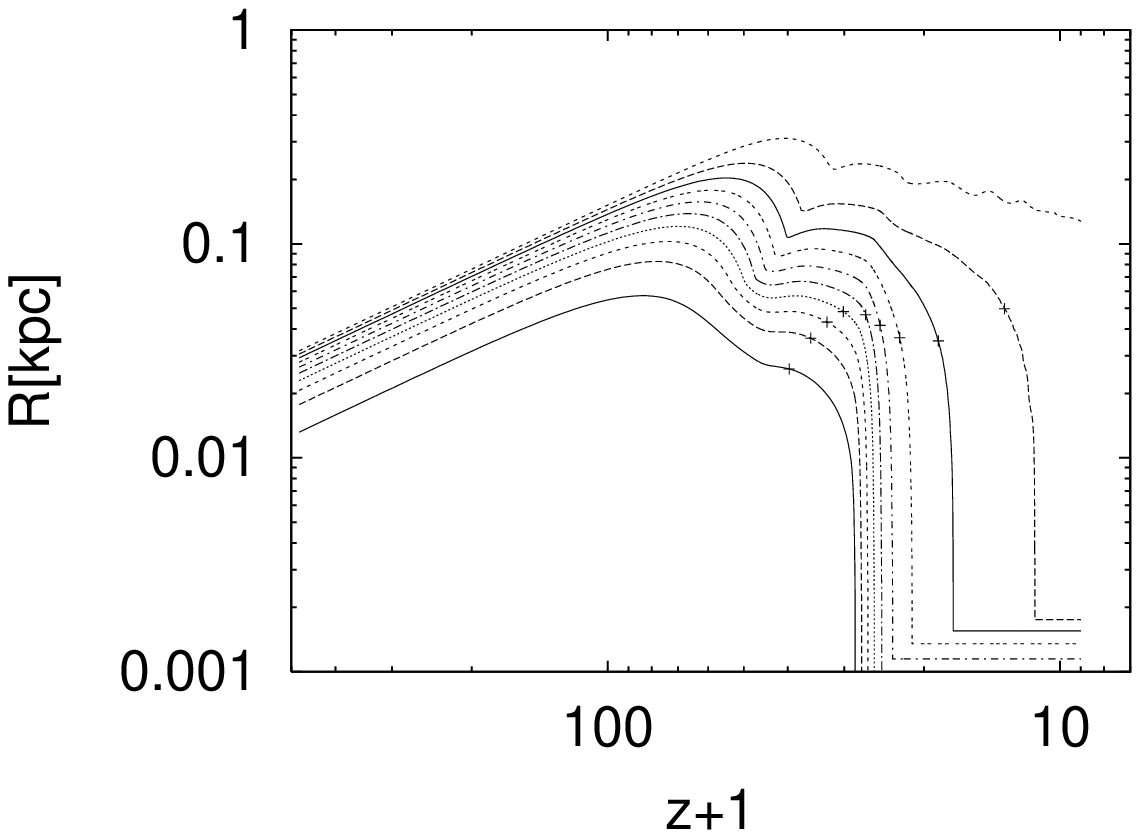}{Shell trajectories for the 'low' $a_0$, $10^{-9}$
overdensity, $M=3\times 10^4 M_\odot$.}{a0m13s3e4}

Most of these figures show trajectories of shells enclosing 7\%, 17\%, 27\% ...
97\% of the total mass: Figs. \ref{a0p01s1e3}-\ref{a0p01s3e4} for the
`standard' $a_0$ and $10^{-9}$ overdensity and  Figs.
\ref{a0m13s3e3}-\ref{a0m13s3e4} for the 'low' $a_0$,
$10^{-9}$ overdensity, respectively. In addition, Fig. \ref{a0p01st} shows
temperatures of the same shells as in Fig. \ref{a0p01s1e3} and Fig.
\ref{a0p01sx} shows
the chemical evolution for shells enclosing $0.12 M_z$, both figures showing
runs for the `standard' $a_0$, $10^{-9}$ overdensity and $M=10^3 M_\odot$ (the
results for the other masses are very similar).
The results for $10^{-8}$ are almost idistinguishable and the run with greater
mass ($M=10^5 M_\odot$) but no H$_2$ cooling showed no collapse.

The results show that:

\begin{itemize}

\item as expected, the difference in behaviour between clouds with $10^{-9}$
and $10^{-8}$
overdensities is very tiny, thus the results are much less sensitive to the
initial density contrast than in CDM

\item more massive clouds collapse faster

\item speed of collapse depends very strongly on $a_0$

\item like in the CDM models, H$_2$ cooling is necessary to collapse

\item like in the CDM models, due to the adiabatic cooling/heating, the shell
temperature behaviour is opposite to that of the shell radii; then H$_2$
cooling becomes
important and the shells collapse; a higher mass of the cloud means higher
virial temperature and faster collapse

\item chemical evolution is quite typical but because the collapse is very
violent, it speeds up chemical reactions; however, the final abundances of
various species are not very different from the predictions of the other
models, e.g. the final abundance of H$_2$ is of the order of $10^{-3}$.

\end{itemize}

It may appear that during the collapse, shell temperatures do not fall as deep
as in CDM models, but this is a numerical artefact only. The reason is, as
in \citet{Tho95}, that in our code -- if a shell falls below some
`small' radius $r_c$ -- we treat it as `collapsed': we artificially stop its
dynamical and chemical evolution, assigning some small radius, freezing the
temperature and including its gravitational field only. A collapse in MOND is
very violent, so gas shells have not enough time to cool down until their
temperature is frozen by the code.

We performed many other runs as well, e.g. with initial overdensities
greater than $10^{-4}$, but the results were very similar.
The reason for this is that, as
shown in the previous section, in the absence of gas pressure, any positive
initial overdensity lower than some reasonable value leads to a very similar
turnaround time. Because at this stage of evolution the gas effects are not yet
very important, this also applies to the evolution with all physical processes
included. This means that in the approximation the turnaround time depends on
mass. Then the cloud starts its initial collapse, but it gets virialised and
the time between the virialisation and the final collapse depends on mass
again, i.e. the bigger the mass, the faster the collapse.

The Wilkinson MAP results suggest that reionization occured at around
$z\sim 20$. This means that the first bound objects must have been formed
even earlier, perhaps at around $z\sim 30$. Our recent simulations
\citep{Sta03} show that if we assume the $\Lambda$CDM models and take the
recent estimates of $\Omega_M$ and $\Omega_\Lambda$, a direct formation of
low-mass objects that could possibly
reionize the Universe before $z\sim 10$ is very unlikely. Moreover, we doubt
whether the inclusion of possible
fragmentation of greater clouds could speed up the collapse enough -- even if
some low-mass cloud has greater overdensity than the directly forming ones, it
still needs some time to cool down.

In contrast, MOND seems to provide a good way to solve that problem. For
the `standard' value of $a_0$ clouds of mass $3\times 10^3 M_{\odot}$ or
heavier may collapse at around $z\sim 30$, so that they or their cores may
have formed the first stars and quasars. For lower $a_0$, only objects of mass
$3\times 10^4 M_{\odot}$ or greater may be formed directly before
$z\sim 30$. We think this favours the `standard' value but one would
need to perform full 3-D simulations to give a more definite answer.

It may be surprising that in our simulations more massive objects form earlier,
but it is due to the gas pressure, which is more important for less massive
objects. If we skip that term in Eq. \ref{predkosc}, less massive objects stop
their expansion and recollapse faster than the more massive ones. However, this
is not the case for the Large Scale Structure: on these scales, gas effects may
be neglected, and although in general the MOND effects are smaller, the rms
fluctuations decrease so that such objects stop their expansion more slowly.

\section{Conclusions}

If our assumptions about MOND are correct, its predictions seem to be
more consistent with the the early reionization suggested by the WMAP results
(\citealt{Ben03}; \citealt{Spe03}) than the ones of the most recent
$\Lambda$CDM models. This does not prove that MOND is correct and $\Lambda$CDM
is not. However, this suggests that cosmologists should perhaps pay more
attention to MOND, because it seems to be an interesting alternative
to models with non-baryonic dark matter.

\section*{Acknowledgments}

We are very grateful to Uro\v{s} Seljak for answering our questions about
{\sc CMBFAST}, to N. Sugiyama for sending us the output of his CMB anisotropy
program, and to R.H. Sanders for his comments about the role of the Silk
damping.

Our work was partially supported by the Polish State Commitee for
Scientific research under grants nrs. 2P03B~11024 and PBZ-KBN-054/P03/02.

\end{document}